\documentclass{article}

\usepackage{lipsum}
\usepackage[T1]{fontenc}
\usepackage[english]{babel}
\usepackage{geometry}
\usepackage{tikz}
\usepackage{booktabs}
\usepackage{float}
\usepackage[english]{varioref}
\usepackage{amsmath,amsfonts,amsthm,amscd,amssymb}
\usepackage{graphicx}
\usepackage{subfigure}
\usepackage{url}
\usepackage{multicol}
\usepackage{hyperref}
\usepackage{authblk}
\usepackage{xcolor}
\usepackage{array}
\usepackage{amssymb}
\usepackage{amsfonts}
\usepackage{graphicx}
\usepackage{enumitem}
\usepackage{amsmath}
\usepackage[round,sort,authoryear]{natbib}

\title{Coevolution of cognition and cooperation in structured populations under reinforcement learning}
\author[1,*]{R. Mastrandrea}
\author[2]{Leonardo Boncinelli}
\author[1]{Ennio Bilancini}

\affil[1]{IMT Alti Studi Lucca, Piazza S. Francesco 19, 55100 Lucca, Italy}
\affil[2]{University of Florence}

\affil[*]{rossana.mastrandrea@imtlucca.it}

\date{}

\begin{document}

\flushbottom
\maketitle

\thispagestyle{empty}

\begin{abstract}
\noindent 
We study the evolution of behavior under reinforcement learning in a Prisoner's Dilemma where agents interact in a regular network and can learn about whether they play one-shot or repeatedly by incurring a cost of deliberation. With respect to other behavioral rules used in the literature, (i) we confirm the existence of a threshold value of the probability of repeated interaction, switching the emergent behavior from intuitive defector to dual-process cooperator; (ii) we find a different role of the node degree, with smaller degrees reducing the evolutionary success of dual-process cooperators; (iii) we observe a higher frequency of deliberation.
\end{abstract}

\section*{Introduction}

The evolution of cooperation has been investigated intensely in various disciplines, such as biology, economics, computer science, physics and psychology. There are two important dimensions, among many \citep{nowak2006five,lehmann2006evolution,bowles2011cooperative}, that have been shown to affect the evolution of cooperation: the interaction structure, i.e., who interacts with whom \citep{santos2006cooperation}, and the mode of cognition, i.e., the extent of deliberation as opposed to intuition \citep{capraro2019dual}. While for the interaction structure there is a substantial consensus that sparse and heavily clustered networks help the spread of cooperation \citep{ohtsuki2006simple,nowak2006five}, for the mode of cognition results are more articulated and depend on specific features of the social dilemma \citep{bear2016intuition,bear2017co} and of the cost of deliberation \citep{jagau2017general}.


An important aspect in evolutionary models is the behavioral rule adopted by agents, which heavily contributes to determining the trajectories of the dynamic adjustment. While the literature has extensively considered behavioral rules encompassing best reply \citep{bilancini2009co} and imitation \citep{levine2007evolution} as well as processes of the type death-birth or birth-death \citep{ohtsuki2006simple}, little attention has been given to evolutionary dynamics based on reinforcement learning \citep{tanabe2012evolution}. Reinforcement learning is a prominent behavioral rule originated in behavioral sciences \citep{skinner1938behavior,skinner1953science} and recently become extremely popular in computer sciences, with many different applications \citep{nian2020review}. Reinforcement learning, which is a particular type of Q-learning, is known for its capacity to solve complex problems, at least those with a fixed environment \citep{watkins1992q}, in a very parsimonious way in terms of informational requirements. However, reinforcement learning performs poorly to attain cooperation in a system with many interacting agents \citep{tanabe2012evolution}. 

We contribute to the literature by addressing the effects of inserting reinforcement learning into the model of Mosleh and Rand \citep{mosleh2018population}, which is basically the same model of Bear and Rand \citep{bear2016intuition} with the addition of an interaction structure in the form of a $k$-regular lattice. In their model, agents accumulate over time payoffs across games played with all of their neighbors, and the agents who are randomly selected to update behavior copy the strategy of a neighbor with a probability that is increasing in the neighbor’s accumulated payoff. In our model, when behavior is updated an agent increases the probability of choices that have best performed in the past, regarding the cognitive mode and the action in the Prisoner's Dilemma conditional on the information acquired.

We provide three main results. First, we confirm previous findings, obtained with other behavioral rules \citep{bear2016intuition, mosleh2018population}, that for low probabilities of repeated interaction the intuitive defector behavior (i.e., always defect, never deliberate) is favored by evolution, while for higher probabilities it is the dual-process cooperator behavior (i.e., cooperate if not deliberating, deliberate to switch to defection when payoff-maximizing) to be favored. The intuition is similar to the one of the previous literature \citep{bear2016intuition,bear2017co, jagau2017general, mosleh2018population}: when the probability of repeated interaction is very low, it is not worth incurring the cost of deliberating to learn if the current interaction is actually repeated or one-shot, with the result that agents always expect the interaction to be mostly one-shot and systematic defection becomes very attractive.
In this respect, reinforcement learning does not change the quality of results.

Second, in contrast with specific \citep{mosleh2018population} and general \citep{ohtsuki2006simple,nowak2006five} findings of the previous literature that fewer connections help the spread of cooperation, we find that a lower degree in the regular network hampers the adoption of dual-process cooperation, making intuitive defection more likely to be adopted. Therefore, reinforcement learning drastically changes the effects of a higher number of connections, which here becomes a factor promoting cooperation.

Third, we find that reinforcement learning increases the observed frequency of deliberation, for every probability of repeated interaction, comparing our results with those of \citet{mosleh2018population}. It can be interesting to notice that a cognitively cheap behavioral rule, as reinforcement learning is, is associated with a higher reliance of the more demanding cognitive mode of deliberation. In turn, this greater reliance on deliberation has an impact on cooperation rates, depending on whether deliberation is more cooperative than intuition (which happens for low probabilities of repeated interaction) or less cooperative (which happens for high probabilities of repeated interaction).

Overall, reinforcement learning does not change the conclusion that there exists a threshold value of the probability of repeated interaction switching the emergent behavior from intuitive defector to dual-process cooperator. At the same time, we can conclude that the role of the number of connections for the spread of cooperation is moderated by the behavioral rule, with a higher network degree favoring cooperation under reinforcement learning. Finally, we observe that reinforcement learning increases the observes frequency of deliberation. 



\section*{Methods}

In this section we introduce the model, the dynamics and simulations setup, highlighting similarities and differences with \cite{mosleh2018population}.
\subsection*{The model}

Our results are based on agent based simulations. Here, we describe the model setup.

\paragraph*{Population.}

The representative population consists of $N = 100$ agents simultaneously playing a modified version of the Prisoner's Dilemma (PD). At each iteration, the two interacting agents can play Tit-for-Tat (TFT)  or always defect (AllD) in a one-shot or
repeated game. In the one-shot game, $TFT$ corresponds to cooperation, while $AllD$ stands for defection. From now on, we will use for simplicity $C$ or $D$ to indicate the two choices in both scenarios. 

\paragraph*{Stage game.}

The game is a repeated PD with probability $P_G$  and payoff matrix 
\[
\begin{bmatrix}
 b-c &  0 \\     
 b & 0 
 \end{bmatrix}
\]

\noindent while with probability $1-p_G$ the agents will be involved in a one-shot PD with payoff matrix

\[
\begin{bmatrix}
 b-c &  -c \\     
 b & 0 
 \end{bmatrix}
 \]
 
 \noindent where $b=4$ and $c=1$ in our simulations.

\paragraph*{Deliberation.} 

At each round, a deliberation cost $d^*$ is randomly drawn from a uniform distribution in $[0,1]$. Each agent has her own individual threshold cost for deliberation, $d_i$, $i \in \{1,\dots,N\}$ : if $d_i \le d*$, the agent will deliberate acquiring information about the game type (one-shot/repeated); if $d_i > d*$, she will act under intuition ignoring the type of game. 

\paragraph*{Interaction network.}

All agents are placed on a regular lattice with fixed number of neighbours for each agent, i.e. fixed \textit{node degree}  $k$. We consider different degree values: $k \in \{2,4,8,20,40\}$. The graph is mathematically represented by an adjacency matrix  $A\equiv(a_{ij})_{1\lq i,j\lq N}$, where $a_{ij}=1$ means that agent $i$ is linked to agent $j$, while $a_{ij}=0$ represents the absence of agents connection.

Each agent is characterized by a vector of parameters, completely defining her strategy:

\begin{equation}\label{strategy}
(p_{i,int},p_{i,del}^{1s},p_{i,del}^{rep}, d_i)   \quad \quad i\in \{1,\dots,N\}
\end{equation}

\medskip

\noindent where $p_{i,int}$ is the probability to play $C$ under intuition (i.e., independently on the type of game); $p_{i,del}^{1s}$ is the probability to play $C$ under deliberation in a one-shot PD; $p_{i,del}^{rep}$ is the probability to play $C$ under deliberation in a repeated PD; $d_i$ is the agent deliberation cost threshold.

At each round, all pairs of connected agents are selected to play the game. The random cost $d^*\sim U[0,1]$ is drawn at the beginning of each round and kept fixed for all playing agents.


\subsection*{Dynamics}


When all agents have played, the \textit{reinforcement learning} rule is applied to update the strategy of each agent, both in terms of actions taken and deliberation cost threshold. At the end of the game, each agent compares the average payoffs of each choice and game-type and decide to increase/decrease the related probabilities accordingly.
We want to stress that each agent can play different strategies (cooperate/defect or decide intuitively/deliberately) with different neighbours in the same round, but according to the same strategy vector  \eqref{strategy} kept fixed. Indeed, the update of the probabilities/cost is done only when all agents have played with all neighbours. 

\paragraph*{Deliberation cost.}

At each time and for each agent, we compute the experiences payoffs of playing under intuition and under deliberation at that time, averaged over the two possible choices $C$ or $D$: $\pi_{i,int}$, $\pi_{i,del}$. Then, the individual deliberation cost is updated according to the following rule:

\begin{equation}\label{cost}
d_{i} = 
\begin{cases}
d_{i} + \lambda&\text{if} \quad \pi_{i,int}<\pi_{i,del}-d_i\\
d_{i} - \lambda&\text{if} \quad \pi_{i,int}>\pi_{i,del}-d_i\\
\end{cases}
\end{equation}

\noindent with $\lambda = 0.1$.

In other terms, the agent increases (decreases) her deliberation cost of a small amount, $\lambda$, if the average payoff obtained under deliberation - net of the deliberation cost - is greater (smaller) than the average payoff obtained acting under intuition. If $\pi_{i,int}=\pi_{i,del}-d_i$, or if one of the two payoffs cannot be calculated, then the agent deliberation cost does not change. 
We set $0$ and $1$ respectively as the maximum and minimum values for the deliberation cost threshold, therefore if $d_i < 0 \rightarrow d_i =0 $ and if $d_i > 1 \rightarrow d_i = 1 $. This choice avoids meaningless outcomes, such as a negative or extremely high deliberation cost threshold, without loss of generality. Indeed, the effect of having a deliberation threshold cost equal to 0 (1) is exactly the same of having a negative (greater than 1) one, but speeding up the process avoiding to reach very small (high) threshold values.
Finally, it is worth to notice that the update is done only when both payoffs are observed (i.e., the agent acts both under deliberation and under intuition at least once in the same round).

\paragraph*{Action strategy.}

At the end of each round, the following average payoffs are computed: (i) the payoffs resulting from cooperation, $\pi_{i,int}^{C}$, and defection, $\pi_{i,int}^{D}$, under intuition; (ii) the payoffs resulting from cooperation, $\pi_{i,del,1s}^{C}$, and  defection, $\pi_{i,del,1s}^{D}$, under deliberation when the game is one-shot; (iii) the payoffs resulting from cooperation, $\pi_{i,del,rep}^{C}$, and defection, $\pi_{i,del,rep}^{D}$, under deliberation when the game is repeated. 

Then, the strategy played by every agent $i$ is updated according to the following rules:

\begin{equation}\label{int}
p_{i,int} = 
\begin{cases}
p_{i,int} + \epsilon&\text{if} \quad \pi_{i,int}^{C}>\pi_{i,int}^{D}\\
p_{i,int} - \epsilon&\text{if} \quad \pi_{i,int}^{C}<\pi_{i,int}^{D}\\
\end{cases}
\end{equation}

\begin{equation}\label{delone}
p_{i,del,1s} = 
\begin{cases}
p_{i,del,1s} + \epsilon&\text{if} \quad \pi_{i,del,1s}^{C}>\pi_{i,del,1s}^{D}\\
p_{i,del,1s} + \epsilon&\text{if} \quad \pi_{i,del,1s}^{C}<\pi_{i,del,1s}^{D}\\
\end{cases}
\end{equation}

\begin{equation}\label{delrep}
p_{i,del,rep} = 
\begin{cases}
p_{i,del,rep} + \epsilon&\text{if} \quad \pi_{i,del,rep}^{C}>\pi_{i,del,rep}^{D}\\
p_{i,del,rep} + \epsilon&\text{if} \quad \pi_{i,del,rep}^{C}<\pi_{i,del,rep}^{D}\\
\end{cases}
\end{equation}

\noindent with $\epsilon = 0.1$.

The probability of cooperating in the three cases (under intuition, under deliberation when the game is one-shot, under deliberation when the game is repeated) is increased (decreased) by a small amount, $\epsilon$, if the related payoff averaged over all games played by the agent in that round (i.e., its number of neighbours or node degree) is greater (smaller) than the average defection payoff obtained respectively in the same three scenarios. 
No update takes place in the above thresholds in case of equality or if one of the payoffs to be compared is absent.

\paragraph*{Perturbations.}

We assume that with probability $\mu = 0.05$ a mutation occurs: the agent updates her action strategy or her deliberation cost exchanging the inequalities in \eqref{cost},\eqref{int},\eqref{delone},\eqref{delrep}. This choice introduce the possibility of a random mistake in the update process at each decision level.

\subsection*{Simulations}

We considered a discrete range of variability for the probability to have a one-shot game with step equal to $0.1$: $p_{G}\in \{0,0.1,0.2,\dots,1\}$. At the beginning of each simulation run, all strategy vectors are independently initialized from  a uniform distribution. Each simulation run continued over a number of generations until no more than one agent updates its strategy for $10^2$ generations. All results are averaged over 1000 initializations. 

\section*{Results}



In this work the same setting proposed by \cite{mosleh2018population} has been employed, except for the behavioural rule used by agents for updating their strategy. Hence, the comparison between our and their outcomes allows us to focus on the effect of reinforcement learning. In the following sections, we highlight three main results standing out from this comparison. 

\paragraph{Emergence of dual-process cooperation.}

\begin{figure}[!ht]
\centering
\subfigure[]
{\includegraphics[width=.48\linewidth]{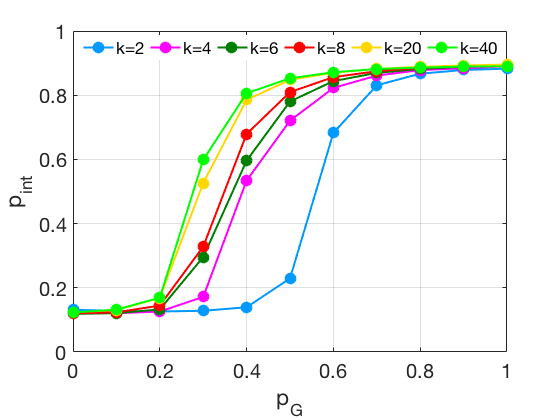}}
\subfigure[]
{\includegraphics[width=.48\linewidth]{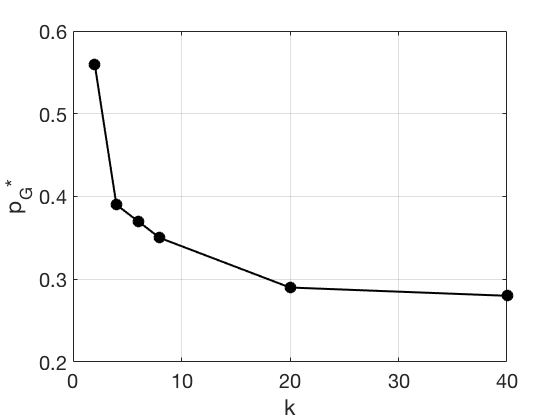}}
\subfigure[]
{\includegraphics[width=.48\linewidth]{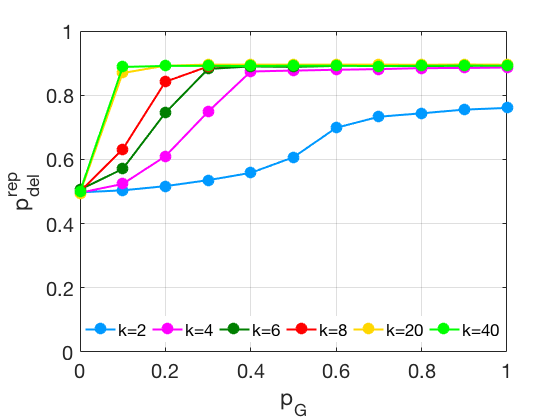}}
\subfigure[]
{\includegraphics[width=.48\linewidth]{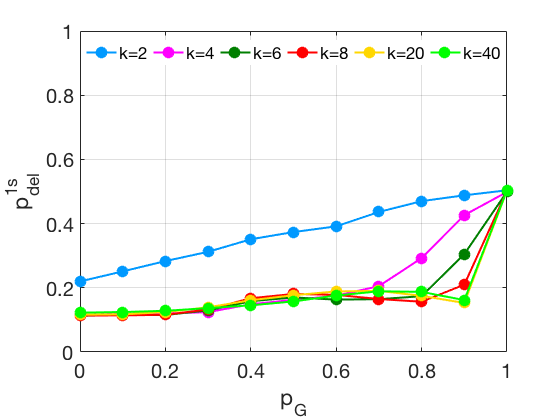}}
\subfigure[]
{\includegraphics[width=.48\linewidth]{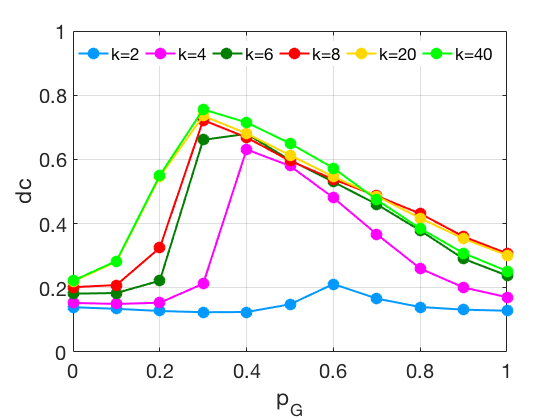}}
\subfigure[]
{\includegraphics[width=.48\linewidth]{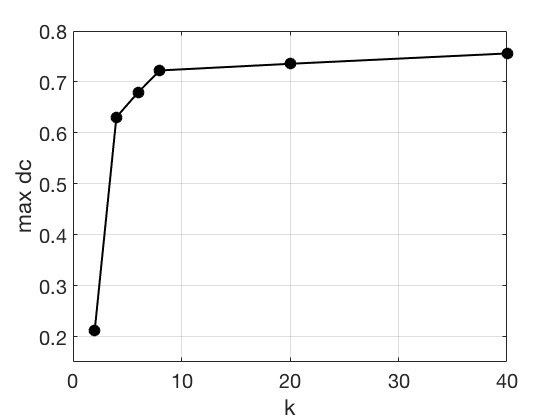}}
\caption{\textbf{Strategy evolution and network structure.} (a) Probability of cooperating under intuition, $p_{int}$, as  function of the probability to have a repeated game, $p_G$. (b) Critical value of the probability that the interaction is repeated, $p_G$, for which $p_{int}=0.5$ across the different values of $k \in \{2,4,6,8,20,40\}$. (c) Probability of cooperating under deliberation when the game is
repeated, $p_{del}^{rep}$, as  function of the probability that the game is repeated, $p_G$. (d) Probability of cooperating under deliberation if the game is one-shot, $p_{del}^{1s}$, as function of the probability that the game is repeated, $p_G$. (e) Maximum threshold cost of deliberation, $d*$, as function of the probability that the game is repeated, $p_G$. (f) Maximum value of the  deliberation cost across the whole discrete range of variability of the probability to have a repeated game, $p_G \in \{0,0.1,\dots,1\}$. In all panels the number of each node's neighbours is fixed,  $k \in\{2,4,6,8,20,40\}$. }
\label{Hom}
\end{figure}

Figure \ref{Hom}(a) shows that as the probability of repeated interaction, $P_G$, increases, the observed frequency of cooperation under intuition gets larger and larger, starting from values close to $0.1$ and reaching almost $1$. These S-shaped curves correspond to the switch from situations mainly characterized by intuitive defection (for low values of $P_G$) to situations where dual-process cooperation prevails (for high values of $P_G$). This is qualitatively the same result as in \citet{mosleh2018population}.

\clearpage

However, a difference emerges for small values of $P_G$: while the observed frequency of cooperation under intuition in \citet{mosleh2018population} is low for large values of $k$ and high for small values of $k$, we find a low frequency of cooperation irrespective of $k$. This outcome suggests that when a reinforcement learning update rule is employed, the network structure/number of neighbours becomes negligible for the emergence of cooperation under intuition when the game is mainly one-shot, remaining its value around $0.1$ independently on the number of games played by each agent.

\paragraph{Dual-process cooperation and node degree.} 
In contrast with Mosleh and Rand, we find that the curves associated with a smaller $k$ are located further down, as shown in figure \ref{Hom}(a). Our results suggest that a higher number of neighbours favors the adoption of dual-process cooperator behavior, while the opposite is true for ~\citet{mosleh2018population}. This outcome highlights the role played by the reinforcement learning update rule: by increasing the number of games played by each agent in each round, the probability to move from intuitive defection to dual-process cooperation rapidly increases; on the contrary for $k=2$ this passage arrives at high value of $p_G$ ($>0.5$). Figure \ref{Hom}(b) summarizes this effect, showing a negative slope curve representing the relation between the node degree and the probability $P_G$ (the analogous figure in \citet{mosleh2018population} exhibits instead a positive slope). 

Furthermore, by looking at figures \ref{Hom}(c) and \ref{Hom}(d), we recognize that the frequency of cooperation under deliberation when the game is repeated rise rapidly as $p_G$ increases, at least when the node degree is rather large, and they never go down. In our case, hence, we do not observe the frequency get pulled back towards $0.5$ as in \citet{mosleh2018population}, figure 3.





\paragraph{Frequency of deliberation.} 

Figure \ref{Hom}(e) shows the maximum threshold cost for deliberation averaged over all agents and for all values of the probability $p_G$. Even if the curves shapes (inverted V) are quite similar to the analogous graph of Mosleh and Rand, in our setting two main differences occur: (i) for all values of $k$, except for $k=2$, we observe very close trends of the maximum deliberating cost with respect to the probability $p_G$ with a slight leftward shift of the peak as $k$ increases (exactly the opposite tendency exhibiting in the analogous figure of Mosleh); (ii) the maximum threshold costs of deliberation double that of Mosleh and Rand for almost all values of $p_G$. Both outcomes, summarised in figure \ref{Hom}(f), suggest a higher frequency of deliberation when a reinforcement learning update rule is used.




\begin{figure}[!ht]
\centering
\subfigure[]
{\includegraphics[width=.48\linewidth]{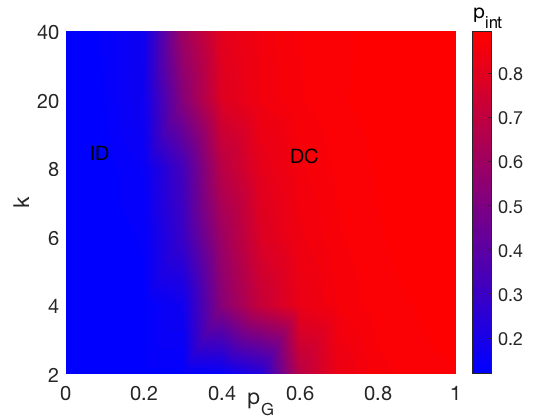}}
\subfigure[]
{\includegraphics[width=.48\linewidth]{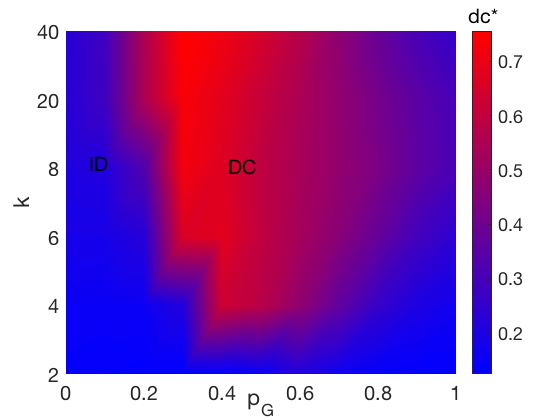}}
\caption{\textbf{Dual-process cooperation and network structure.} Heat-maps of the (a) probability of cooperating under intuition, $p_{int}$, and (b) the maximum threshold cost of deliberation, $d^*$, as a function of the number of neighbours (i.e., the number of games played by each agent in each round), $k$, and the probability to have a repeated game, $p_G$.  }
\label{HomHM}
\end{figure}

The heat-maps in figure \ref{HomHM} provide a graphical summary of our main results showing the variation of the probability of cooperating under intuition, in figure \ref{HomHM}(a), and the maximum threshold cost, in figure \ref{HomHM} (b), with respect to the node degree, $k$, and the probability to have a repeated game, $p_G$. This is particularly useful to highlight the differences with the setup proposed by Mosleh \cite{mosleh2018population}, hence, to evaluate the effect of a reinforcement learning update rule.

Finally, we have repeated the same analysis with different network structures: Erd\"os-R\'enyi random graph, Watt-Strogatz and Barabasi-Albert model with varying node degree ($k \in\{2,,6,8,20,40\}$). These findings show high similarity to the case of agents placed on regular lattices, (i) confirming that the observed outcomes are related to the network density rather then to other specific network features as already pointed out by Mosleh and Rand; (ii) stressing that the differences observed with respect to their setup  are due to the introduction of a reinforcement learning update rule (figures available upon request).


\section*{Discussion}

The debate in the literature whether people are more cooperative under intuition or deliberation has not focused much so far on the specific rule adopted by agents when updating behavior. In this paper we have shown that the main qualitative conclusion obtained in \citet{mosleh2018population} is robust to agents following a reinforcement learning rule: for a low probability of repeated interaction the prevailing behavior is intuitive defection, while dual-process cooperation spreads as such probability increases until it becomes prevalent for high probabilities. This pattern holds for every number of interacting partners. In this respect we differ from \citet{mosleh2018population}, where dual-process cooperation is quite common also for low probabilities of repeated interaction when the number of interacting partners is small. Figure 4 summarizes our main findings.

At the same time, there are important qualitative differences brought about by reinforcement learning.   
The most important one concerns the widespread belief in literature that a smaller number of interacting partners promotes the spread of cooperation. With reinforcement learning, we find the opposite: the smaller the number of partners, the more difficult it is for the dual-process cooperation to prevail. The reasons for this result could be fruitfully investigated in future research.

Another interesting observation concerns the observed frequency of deliberation. The use of reinforcement learning, which is a behavioral rule that consumes few cognitive resources, increases the frequency of deliberation, which is more cognitively costly than intuition.

In general, the results of our study call for further investigation into the role played by behavioral rules in explaining the prevalence of cooperation and intuition within models of dual-process theory. Along this research line, it would be interesting to explore cases where the behavioral rules differs between deliberation and intuition, while in this paper, and in related studies in the literature, only the available information changes between one cognitive mode and the other. 

Finally, it is also interesting to study how the network of interactions evolves over time in response to the payoffs earned. This would require a model where cognition and cooperation coevolve over a dynamic network, which we leave for further research.

\subsection*{Acknowledgments}
We also gratefully acknowledge financial support from the Italian Ministry of Education, University and Research (MIUR) through the PRIN project Co.S.Mo.Pro.Be.~``Cognition, Social Motives and Prosocial Behavior'' (grant n.~20178293XT) and from the IMT School for Advanced Studies Lucca through the PAI project Pro.Co.P.E.~``Prosociality, Cognition, and Peer Effects''.

\bibliographystyle{agsm}

\bibliography{biboma}

@article{bear2016intuition,
  title={Intuition, deliberation, and the evolution of cooperation},
  author={Bear, Adam and Rand, David G},
  journal={Proceedings of the National Academy of Sciences},
  volume={113},
  number={4},
  pages={936--941},
  year={2016},
  publisher={National Acad Sciences}
}

@article{bear2017co,
  title={Co-evolution of cooperation and cognition: the impact of imperfect deliberation and context-sensitive intuition},
  author={Bear, Adam and Kagan, Ari and Rand, David G},
  journal={Proceedings of the Royal Society B},
  volume={284},
  number={1851},
  pages={20162326},
  year={2017},
  organization={The Royal Society}
}

@article{nowak2006five,
  title={Five rules for the evolution of cooperation},
  author={Nowak, Martin A},
  journal={Science},
  volume={314},
  number={5805},
  pages={1560--1563},
  year={2006},
  publisher={American Association for the Advancement of Science}
}

@book{bowles2011cooperative,
  title={A Cooperative Species: Human Reciprocity and its Evolution},
  author={Bowles, Samuel and Gintis, Herbert},
  year={2011},
  publisher={Princeton University Press}
}

@article{jagau2017general,
  title={A general evolutionary framework for the role of intuition and deliberation in cooperation},
  author={Jagau, Stephan and van Veelen, Matthijs},
  journal={Nature Human Behaviour},
  volume={1},
  number={8},
  pages={1--6},
  year={2017},
  publisher={Nature Publishing Group}
}

@article{mosleh2018population,
  title={Population structure promotes the evolution of intuitive cooperation and inhibits deliberation},
  author={Mosleh, Mohsen and Rand, David G},
  journal={Scientific Reports},
  volume={8},
  number={1},
  pages={1--8},
  year={2018},
  publisher={Nature Publishing Group}
}

@article{capraro2019dual,
  title={The dual-process approach to human sociality: A review},
  author={Capraro, Valerio},
  journal={Available at SSRN 3409146},
  year={2019}
}

@article{ohtsuki2006simple,
  title={A simple rule for the evolution of cooperation on graphs and social networks},
  author={Ohtsuki, Hisashi and Hauert, Christoph and Lieberman, Erez and Nowak, Martin A},
  journal={Nature},
  volume={441},
  number={7092},
  pages={502--505},
  year={2006},
  publisher={Nature Publishing Group}
}

@article{bilancini2009co,
  title={The co-evolution of cooperation and defection under local interaction and endogenous network formation},
  author={Bilancini, Ennio and Boncinelli, Leonardo},
  journal={Journal of Economic Behavior \& Organization},
  volume={70},
  number={1-2},
  pages={186--195},
  year={2009},
  publisher={Elsevier}
}

@article{lehmann2006evolution,
  title={The evolution of cooperation and altruism--a general framework and a classification of models},
  author={Lehmann, Laurent and Keller, Laurent},
  journal={Journal of Evolutionary Biology},
  volume={19},
  number={5},
  pages={1365--1376},
  year={2006},
  publisher={Wiley Online Library}
}

@article{levine2007evolution,
  title={The evolution of cooperation through imitation},
  author={Levine, David K and Pesendorfer, Wolfgang},
  journal={Games and Economic Behavior},
  volume={58},
  number={2},
  pages={293--315},
  year={2007},
  publisher={Elsevier}
}

@article{tanabe2012evolution,
  title={Evolution of cooperation facilitated by reinforcement learning with adaptive aspiration levels},
  author={Tanabe, Shoma and Masuda, Naoki},
  journal={Journal of Theoretical Biology},
  volume={293},
  pages={151--160},
  year={2012},
  publisher={Elsevier}
}

@article{santos2006cooperation,
  title={Cooperation prevails when individuals adjust their social ties},
  author={Santos, Francisco C and Pacheco, Jorge M and Lenaerts, Tom},
  journal={PLoS Computational Biology},
  volume={2},
  number={10},
  pages={e140},
  year={2006},
  publisher={Public Library of Science San Francisco, USA}
}

@article{nian2020review,
  title={A review on reinforcement learning: Introduction and applications in industrial process control},
  author={Nian, Rui and Liu, Jinfeng and Huang, Biao},
  journal={Computers \& Chemical Engineering},
  volume={139},
  pages={106886},
  year={2020},
  publisher={Elsevier}
}

@book{skinner1938behavior,
  title={The behavior of organisms: An experimental analysis},
  author={Skinner, Burrhus Frederic},
  year={1938},
  publisher={Appleton-Century-Crofts}
}

@book{skinner1953science,
  title={Science and human behavior},
  author={Skinner, Burrhus Frederic},
  year={1938},
  publisher={Macmillan}
}

@article{watkins1992q,
  title={Q-learning},
  author={Watkins, Christopher JCH and Dayan, Peter},
  journal={Machine Learning},
  volume={8},
  pages={279--292},
  year={1992},
  publisher={Springer}
}

\end{document}